\documentclass[lettersize,journal]{IEEEtran}

\usepackage{cite}
\usepackage{amsmath,amsfonts,amssymb}
\usepackage{algorithmic}
\usepackage{array}
\usepackage[caption=false,font=normalsize,labelfont=sf,textfont=sf]{subfig}
\usepackage{textcomp}
\usepackage{stfloats}
\usepackage{url}
\usepackage{verbatim}

\usepackage{graphicx}
\usepackage{xcolor}
\def\BibTeX{{\rm B\kern-.05em{\sc i\kern-.025em b}\kern-.08em
    T\kern-.1667em\lower.7ex\hbox{E}\kern-.125emX}}
\usepackage{balance}

\begin{document}
\title{Radiation Response of NbN Superconducting Nanowire Single-Photon Detectors Under $^{210}$Po Alpha-Particle Irradiation}

\author{Francesca~Incalza\textsuperscript{},
        Matteo~Castellani,
        Alejandro~Simon,
        Emma~Batson,
        and Karl~K.~Berggren%
\thanks{This work was supported by an Early Stage Innovations grant from NASA’s Space Technology Research Grants Program (80NSSC24K0277). The work of A. S. and E. B. was supported by the National Science Foundation Graduate Research Fellowship Program (NSF GRFP).}
\thanks{The authors are with the Department of Electrical Engineering and Computer Science, Massachusetts Institute of Technology, Cambridge, MA 02139 USA.}%
\thanks{Corresponding author: Francesca Incalza (e-mail: incal723@mit.edu).}}

\markboth{}%
{Incalza \MakeLowercase{\textit{et al.}}: Radiation Response of NbN Superconducting Nanowire Single-Photon Detectors Under $\boldsymbol{^{210}\mathrm{Po}}$ Alpha-Particle Irradiation}

\IEEEaftertitletext{\vspace{-2.2\baselineskip}}

\maketitle

\begin{abstract}
%\doublespacing

Superconducting nanowire single-photon detectors (SNSPDs) are attractive for space applications because they combine high detection efficiency, low noise, and fast timing, but their radiation tolerance remains less established than that of conventional semiconductor detectors. Here we study NbN SNSPDs exposed to alpha-particle Po-210 irradiation by combining cryogenic electrical and photon-counting measurements with SRIM and Geant4 radiation-transport calculations. For the estimated fluence corresponding to the 200-hour $\boldsymbol{^{210}\mathrm{Po}}$ exposure, the air-inclusive Geant4 calculation predicts an average ionizing dose of 8.49~Mrad in the active NbN layer and a displacement-damage estimate of $\boldsymbol{5.67\times10^{-7}}$ displacements per atom. Measurements before and after irradiation show no systematic degradation of the superconducting transition, switching current, retrapping current, dark-count behavior, or
photon-count response. These results indicate radiation resilience of
thin-film NbN SNSPDs under a multi-Mrad alpha-particle exposure and
support their use in photon-counting systems for radiation environments.
\end{abstract}

%\doublespacing

\section{Introduction}
\IEEEPARstart{S}{uperconducting} nanowire single-photon detectors (SNSPDs) enable quantum communication~\cite{colangelo2024molybdenum, hadfield2009single}, dark matter detection~\cite{chiles2022new, pena2025high}, and deep-space communication~\cite{ivanov2018design, wollman2021recent} because they offer low dark-count rates~\cite{shibata2015ultimate}, high efficiency~\cite{reddy2020superconducting, ma2025drone}, and sub-nanosecond timing response~\cite{korzh2020demonstration}. In space and accelerator-adjacent environments, these advantages must be maintained under radiation exposure~\cite{anisimova2017mitigating}. Radiation tolerance is also relevant beyond single-photon detection; superconducting electronics have long been considered for cryogenic digital systems in space environments because of their high speed and low dissipation~\cite{harris1980space,pagano1997radiation, medeiros2026scalable, guay2026reconfigurable}. 
Near-Earth radiation environments include trapped and solar protons and electrons, as well as helium nuclei and heavier ions associated with solar energetic particles and galactic cosmic rays; the particle spectra and fluences incident on a device depend on orbit, mission duration, solar activity, and shielding~\cite{xapsos2012near}. These particles can affect device performance through two cumulative radiation mechanisms, total ionizing dose (TID) and displacement damage, and through single event effects caused by individual particle interactions during device operation~\cite{xapsos2012near, rius2026radiation}. Accordingly, both cumulative irradiation effects and transient particle responses have been investigated in MKIDs/KIDs, superconducting microwave resonators, TESs, and SQUIDs~\cite{karatsu2019mitigation,kane2026radiation,
besnard2026effect,thobaben2026effect,osherson2020particle,
naruse2020niobium,casagranda2026squid, vepsalainen2020impact}. For superconducting nanowires, one of the most relevant long-term questions is whether ionization or lattice disorder measurably degrades the superconducting transition, switching current, or photon-counting characteristics.
\begin{figure}[h!]
    \centering
    \includegraphics[width=\linewidth]{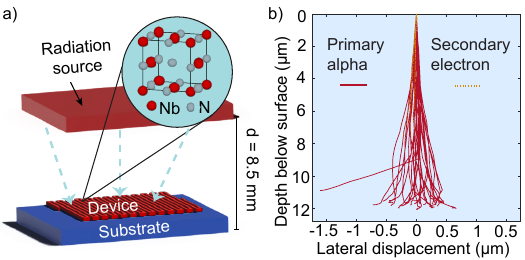}
    \caption{Irradiation geometry and simulated material stack. 
(a) Schematic of the $^{210}$Po alpha-particle irradiation experiment. NbN SNSPDs were exposed at room temperature for 200~h across an
8.5-mm air gap and subsequently remeasured under the same cryogenic
conditions. The source emits 5.304-MeV alpha particles, which reach
the sample surface with a simulated mean energy of 4.513~MeV.
(b) Representative particle trajectories calculated using Geant4. Normally incident 5.304-MeV alpha particles were transported through 8.5 mm of air before entering the 7-nm HSQ/5-nm NbN/Al\(_2\)O\(_3\) stack, reaching the HSQ surface with a mean energy of approximately 4.513 MeV. Thirty representative primary-alpha trajectories from a 200-particle run are shown as solid red lines, together with selected secondary-electron trajectories with energies above 1 keV, shown as dotted orange lines. The trajectories are projections of recorded Geant4 step endpoints onto the lateral–depth plane without smoothing or fitting. Depth is measured from the top HSQ surface, and, for each event, lateral displacement is referenced to the primary alpha particle’s entry position. The air gap is included in the transport calculation but is not displayed.}
\label{fig:setup}
    \label{fig:setup}
\end{figure}
Radiation studies of single-photon detectors have largely focused on semiconductor technologies, for which displacement damage often increases dark-count rates and requires mitigation by cooling or annealing~\cite{anisimova2017mitigating}. Superconducting thin films provide a different materials platform: they do not rely on carrier doping in the same way as semiconductor junctions, their nanoscale active layers can limit the amount of displacement damage accumulated within the superconducting film, and prior work has shown that controlled disorder can sometimes be compatible with, or even beneficial for, nanowire detector operation~\cite{zhang2019saturating, charaev2024single, batson2024effects, incalza2023effects, hong2025impact, strohauer2025origin}. Related particle-interaction studies have shown that NbTiN superconducting nanowires can efficiently detect alpha and beta particles while exhibiting negligible response to gamma rays~\cite{azzouz2012efficient}. More recently, NbN superconducting nanowires have also been used for direct detection of 120-GeV protons~\cite{lee2024beam}. These experiments demonstrate sensitivity to individual energetic particles but do not establish the cumulative radiation tolerance of SNSPDs. In niobium nitride (NbN) thin films, neutron irradiation to a fluence of $10^{23}$~m$^{-2}$, corresponding to an estimated displacement probability of 0.009 displacements per atom,
reduced the film critical temperature by only 5-7\% while leaving the
critical-current density unchanged~\cite{gregshammer1988effects}. However, the radiation response of 5-nm-thick NbN SNSPDs remains insufficiently quantified for space-relevant photon-counting applications.

In this work, we focus on the total ionizing dose (TID) and displacement damage by combining irradiation experiments with depth-resolved radiation-transport simulations. Two NbN SNSPDs were characterized before and after alpha-particle irradiation to assess changes in their superconducting and photon-counting properties. Two nominally identical unirradiated devices were characterized over the same time interval and used as controls to provide a reference for measurement variability. As illustrated in Fig.~\ref{fig:setup}(a), the irradiated devices were exposed to alpha particles from a $^{210}$Po source, while the control devices were exposed only to ambient conditions. In parallel, SRIM~\cite{ziegler2010srim} and Geant4~\cite{agostinelli2003geant4, asai2015recent} calculations were used to estimate the ionizing dose and displacement damage generated in the HSQ/NbN/Al$_2$O$_3$ stack, with the 5~nm depth interval corresponding to the active 5-nm NbN layer, shown in Fig.~\ref{fig:setup}(b) at the experimentally estimated particle fluence which is interpreted as a controlled, accelerated laboratory exposure. This combined experimental and computational approach connects the observed device-level stability to the microscopic energy-deposition and damage profiles in the active NbN layer.

\section{Methods}

In this section, we describe device fabrication and cryogenic measurements,
the alpha-particle irradiation and fluence estimate, and the SRIM and
Geant4 simulations used to calculate ionizing dose and displacement
damage.

\subsection{Device Fabrication and Measurement}
\begin{figure}[!t]
    \centering
    \includegraphics[width=\linewidth]{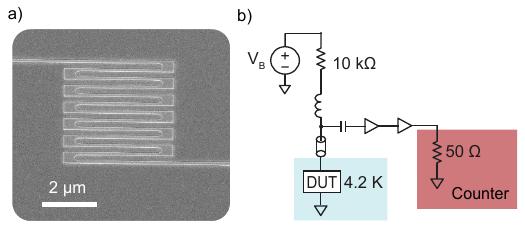}
    %\caption{Device and measurement configuration. (a) Scanning electron micrograph of a representative NbN nanowire detector. (b) Simplified bias and readout circuit used for cryogenic transport and counting measurements.}
    \caption{Device geometry and cryogenic measurement configuration.
    (a) Scanning electron micrograph of a representative NbN SNSPD used in this
    study. The devices consist of 5-nm-thick NbN nanowire meanders patterned on
    sapphire, with a $4~\mu\mathrm{m}\times4~\mu\mathrm{m}$ active area and a
    nominal nanowire width of 100 nm.
    (b) Simplified schematic of the electrical bias and readout configuration used
    for cryogenic transport and photon-counting measurements. The device under test
    (DUT) is operated in liquid helium at 4.2 K and biased using a low-noise voltage
    source through a 10~k$\Omega$ series resistor. The detector output is coupled
    through the readout chain to a 50~$\Omega$ load and counter for pulse detection
    and counting. Photon-counting measurements were performed under continuous-wave
    780-nm optical illumination delivered to the device through an optical fiber.}
    \label{fig:device}
\end{figure}

NbN SNSPDs were patterned from 5-nm-thick NbN films on sapphire (Al$_2$O$_3$) using high-resolution electron-beam lithography with 2\% HSQ negative-tone resist developed in 25\% tetramethylammonium hydroxide (TMAH), and
a CF$_4$-based dry etch to transfer the pattern into the NbN layer
~\cite{incalza2025fast}. After pattern transfer, approximately 7~nm of HSQ remained on top of the NbN. The devices are nanowire meanders with a 4\,\textmu m $\times$ 4\,\textmu m active area, 100-nm wire width, a fill factor of approximately 33\% and a pre-fabrication film critical temperature of approximately 12~K (Fig.~\ref{fig:device}(a)).

Electrical and photon-counting measurements were performed using a custom-built dip probe immersed in liquid helium at 4.2~K~\cite{butters2022digital} before and after irradiation. The devices were biased through a low-noise battery and  a $10\,\mathrm{k}\Omega$ series resistor. The output pulses were AC-coupled to the amplification and acquisition chain through a bias tee (Fig.~\ref{fig:device}(b)). Optical measurements used continuous-wave 780~nm illumination delivered through a fiber.
For each bias sweep, the light-count response was calculated as the light-count rate (LCR) minus the dark-count rate (DCR),
$\mathrm{LCR}-\mathrm{DCR}$, and normalized to the maximum value of that sweep. To quantify irradiation-induced changes in the photon-count, we extracted the normalized bias current required to reach a
fraction $p$ of the normalized photon-count response, with
$p=0.2$, 0.5, and 0.7 before and after the irradiation. The reported shift was defined as $\Delta(I_b/I_c)_p =
(I_{b,p}/I_c)_{\mathrm{after}}-
(I_{b,p}/I_c)_{\mathrm{before}}$,
where $I_{b,p}$ is the bias current at which the normalized photon-count
response reaches $p$. The DCR bias dependence was quantified by fitting $\log_{10}(\mathrm{DCR})$ as a function of bias current over the range $10^3 \leq \mathrm{DCR} \leq 10^5$ counts/s and comparing the fitted slopes before and after exposure. The superconducting transition temperature $T_\mathrm{c}$, transition width $\Delta T_\mathrm{c}$, switching current $I_\mathrm{sw}$, retrapping current $I_\mathrm{r}$, dark-count rate (DCR), and photon-count response were measured before and after irradiation.

\subsection{Alpha-Particle Irradiation and Fluence Estimate}

Alpha-particle irradiation was performed at room temperature using a commercial $^{210}$Po source (PO-210-500, Images Scientific Instruments, Inc.) with approximate dimensions of $22.2\,\mathrm{mm}\times25.4\,\mathrm{mm}\times6.4\,\mathrm{mm}$, nominal alpha energy of 5.304~MeV, initial activity of 500~$\mu$Ci, and half-life of approximately 138~days~\cite{imagesco_po210_source}. The source included Au and Ag foils above the active $^{210}$Po layer, and the devices were irradiated across an
8.5-mm air gap. The Au and Ag source-encapsulation foils were not modeled explicitly. In Geant4, 5.304-MeV alpha particles were launched at the upstream boundary of the 8.5-mm air gap and transported through air before entering the HSQ/NbN/Al$_2$O$_3$ stack. The simulated alpha particles reached the HSQ surface with a mean energy of 4.513 MeV. In SRIM, the air-induced energy loss was represented by launching 4.513-MeV alpha particles directly at the HSQ surface, thereby retaining depth resolution in the solid stack.
The original $10\,\mathrm{mm}\times10\,\mathrm{mm}$ chip, containing four NbN SNSPDs, was diced in half into two $5\,\mathrm{mm}\times10\,\mathrm{mm}$ chips, each containing two devices. All four devices were characterized before irradiation. Two devices were placed under the alpha-particle source for
a 200~hour exposure. Two nominally identical devices were kept as unirradiated controls under ambient laboratory conditions over the same interval.
After exposure, all devices were re-measured at 4.2~K under identical cryogenic conditions. The controls were used to estimate measurement repeatability and variability from handling, cooldown, thermal cycling, and long-term characterization.
The alpha-particle fluence was estimated from source activity, radioactive decay, source-to-sample geometry, and exposure duration. The active source area was modeled as $A_s=1~\mathrm{cm}\times2~\mathrm{cm}$, based on a visual estimate of the source window dimensions, and located at $d=0.85$~cm from the sample. Assuming emission into the forward hemisphere, the geometric collection factor was
\begin{equation}
f_{\Omega}
=
\frac{1}{2\pi}
\int_{A_s}
\frac{d}
{\left(x^2+y^2+d^2\right)^{3/2}}
\,dA .
\label{eq:geom_factor}
\end{equation}
The local fluence was calculated as $\Phi_{\alpha}=(A_0/A_s)\exp(-\lambda t_{\mathrm{elapsed}})f_{\Omega}[1-\exp(-\lambda t_{\mathrm{exp}})]/\lambda$, where $\lambda=\ln(2)/T_{1/2}$. Using the experimental parameters gives an estimated local alpha-particle fluence of $\Phi_{\alpha}=1.27\times10^{12}~\mathrm{alpha/cm^2}$ at the sample. Across the $4~\mu\mathrm{m}\times4~\mu\mathrm{m}$ device footprint, this fluence corresponds to approximately $2.03\times10^5$ incident alpha particles. The Monte Carlo transport results were normalized per incident alpha particle and subsequently scaled to the experimental fluence.

\subsection{Radiation-Transport Simulations}

SRIM and Geant4 simulations were used to estimate depth-dependent ionizing dose and displacement damage in the HSQ/NbN/Al$_2$O$_3$
stack~\cite{ziegler2010srim,agostinelli2003geant4,asai2015recent}.
Total ionizing dose (TID) quantifies the ionizing energy deposited per unit mass, whereas displacement damage is expressed as displacements per atom (DPA), which quantifies the average number of atomic displacements per atom. The simulated stack consisted of 8.5 mm of air followed by a 7-nm solid HSQ cap, modeled using the stoichiometric repeat unit H$_2$Si$_2$O$_3$ and a mass density of $1.398~\mathrm{g/cm^3}$; a 5-nm stoichiometric NbN film with a mass density of $8.45~\mathrm{g/cm^3}$; and a 40-$\mu$m sapphire substrate, modeled as homogeneous, isotropic Al$_2$O$_3$ with a mass density of $3.97~\mathrm{g/cm^3}$. The SRIM calculation used the same solid-stack composition and dimensions but began at the HSQ surface with an incident alpha-particle energy of 4.513 MeV. This approximation is not expected to materially affect the TID or DPA in the upstream NbN layer. The air was represented by \texttt{G4\_AIR} with a density of $1.20479\times10^{-3}$~g/cm$^3$. The NbN layer was used for the device-relevant dose and displacement-damage extraction.

\paragraph{Ionizing dose}
The total ionizing dose (TID) was calculated from the electronic
energy deposited per unit mass. Here $z$ denotes depth below the
sample surface, $\Phi_\alpha$ is the alpha-particle fluence, and
$\rho(z)$ is the material mass density. In SRIM, the ionizing stopping profile was taken as
$S_{\mathrm{ion}}(z)=S_{\mathrm{ion},p}(z)+S_{\mathrm{ion},r}(z)$,
where $S_{\mathrm{ion},p}$ and $S_{\mathrm{ion},r}$ are the electronic
energy-loss contributions from primary ions and recoils, respectively.
The SRIM stopping profile was converted to dose using
$D_{\mathrm{SRIM}}(z)=
\Phi_\alpha S_{\mathrm{ion}}(z)10^8
(1.602\times10^{-16})/\rho(z)$,
where $S_{\mathrm{ion}}$ is in eV/(\AA$\cdot$ion), $\rho$ is in
g/cm$^3$, the factor $10^8$ converts \AA$^{-1}$ to cm$^{-1}$, and
$1.602\times10^{-16}$ converts eV/g to Gy.

In Geant4, 5.304-MeV alpha particles were transported through 8.5 mm of air before entering the HSQ/NbN/Al\(_2\)O\(_3\) stack. The mean alpha-particle energy at the HSQ surface was 4.513 MeV. Energy deposition was scored in 0.5-nm-thick depth bins, such that the 5-nm NbN layer was represented by 10 depth bins. For each bin, $\overline{E}_{\mathrm{dep}}(z)$ is the mean total deposited energy per incident alpha particle and $\overline{E}_{\mathrm{NIEL}}(z)$ is the corresponding non-ionizing energy loss. The ionizing contribution was calculated as
$E_{\mathrm{ion}}(z)=
\max[E_{\mathrm{dep}}(z)-E_{\mathrm{NIEL}}(z),0]$.
The fluence-scaled dose was then calculated as
$D_{\mathrm{G4}}(z)=
E_{\mathrm{ion}}(z)\Phi_\alpha/
[\rho(z)\Delta z]$,
where $\rho(z)\Delta z$ is the material mass per unit area
in the depth bin. The appropriate energy-to-dose conversion was applied using $1~\mathrm{eV}=1.602\times10^{-19}~\mathrm{J}$ and $1~\mathrm{Mrad}=10^4~\mathrm{Gy}$. The model treats the NbN as a continuous planar film and does not explicitly include the nanowire-meander fill factor or lateral edges.

\paragraph{Displacement damage}

Displacement damage was quantified as displacements per atom (DPA).
Here $n_{\mathrm{at}}(z)$ denotes the atomic number density of the
material at depth $z$. In SRIM, the vacancy profile was taken as
$v(z)=v_{\mathrm{ion}}(z)+v_{\mathrm{recoil}}(z)$, where
$v_{\mathrm{ion}}(z)$ and $v_{\mathrm{recoil}}(z)$ are the vacancy
production profiles due to the incident ions and recoiling target atoms,
respectively. The vacancy profile was converted to
$\mathrm{DPA}_{\mathrm{SRIM}}(z)=
\Phi_\alpha v(z)10^8/n_{\mathrm{at}}(z)$,
where $v(z)$ is in vacancies/(\AA$\cdot$ion), and the factor $10^8$
converts \AA$^{-1}$ to cm$^{-1}$.

\begin{figure}[!t]
    \centering
    \includegraphics[width=\linewidth]{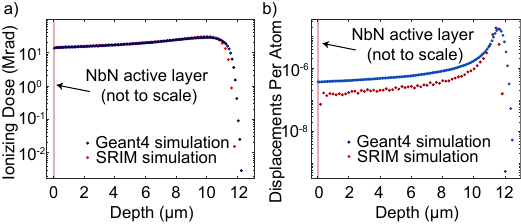}
    \caption{Simulated radiation-damage profiles in the
    HSQ/NbN/Al$_2$O$_3$ stack. In Geant4, alpha particles were
    emitted with an energy of 5.304~MeV and transported through an
    8.5-mm air gap before reaching the HSQ surface. In SRIM, the
    air-induced energy loss was represented using an incident
    energy of 4.513~MeV at the HSQ surface. Depth is measured from
    the top HSQ surface, and the air layer is not shown on the
    horizontal axis. (a) Ionizing-dose profile. (b) Corresponding
    displacement-damage profile. The lower first SRIM
    displacement-damage point is an entrance-bin effect: the
    approximately 0.20-$\mu$m SRIM depth sampling does not resolve
    the 7-nm HSQ and 5-nm NbN layers, and the first bin is also
    affected by recoil escape at the entrance surface. The active NbN layer is located near the sample surface, whereas the maximum displacement damage occurs
    deeper in sapphire near the alpha-particle end of range.}    
    \label{fig:simulation}
\end{figure}

In Geant4, the depth-resolved NIEL was converted to a Norgett--Robinson--Torrens (NRT) displacement estimate~\cite{norgett1975proposed}. For each depth bin, the mean number of NRT displacements produced per incident alpha particle was calculated as $\overline{N}_{\mathrm{disp}}(z)=
0.8E_{\mathrm{NIEL}}(z)/[2E_d(z)]$.
The fluence-scaled DPA was then calculated as $\mathrm{DPA}_{\mathrm{G4}}(z)
= \overline{N}_{\mathrm{disp}}(z)\Phi_{\alpha}/
[n_{\mathrm{at}}(z)\Delta z]$, where $n_{\mathrm{at}}(z)$ is the atomic number density and $\Delta z$ is the depth-bin thickness. For the device-relevant NbN layer, we adopted an effective baseline displacement threshold energy of $E_d=30~\mathrm{eV}$ and evaluated the sensitivity of the result over the range 20-35~eV. The layer-averaged NbN DPA was calculated as the average of the local dimensionless DPA values over the NbN thickness,
\begin{equation}
\left<\mathrm{DPA}_{\mathrm{NbN}}\right>
=
\frac{1}{t_{\mathrm{NbN}}}
\int_{z_1}^{z_2}\mathrm{DPA}(z)\,dz ,
\label{eq:nbn_average_dpa}
\end{equation}
where $t_{\mathrm{NbN}}=5$~nm. Because the NbN layer was modeled with uniform composition and atomic number density and was divided into equal-thickness depth bins, this thickness average is equivalent to the total number of NRT displacements divided by the total number of NbN atoms. Because SRIM determines displacement damage from vacancy production whereas Geant4 uses an NRT conversion of the non-ionizing energy loss, the calculated DPA magnitudes are model dependent. The comparison is therefore used primarily to identify robust depth-dependent trends rather than to require quantitative numerical agreement.

\section{Results and Discussion}

We compare the simulated radiation-damage profiles and the superconducting and photon-counting responses of irradiated and control devices.

\subsection{Simulation Results}

We simulated ionizing-dose and displacement-damage profiles for the 5.304-MeV alpha-particle exposure. The simulated TID and DPA profiles are shown in Fig.~\ref{fig:simulation}(a) and Fig.~\ref{fig:simulation}(b), respectively. Both SRIM and Geant4 place the active NbN layer in the entrance region
of the alpha-particle trajectory and predict that the displacement-damage
maximum occurs deeper in the sapphire substrate near the alpha-particle
end of range. The two calculations show the same overall qualitative
depth dependence, although quantitative differences are expected because
SRIM and Geant4 use different stopping and displacement-damage formalisms.
Averaging over the 5-nm NbN layer gives a fluence-scaled Geant4 ionizing
dose of $\left<D_{\mathrm{TID,NbN}}\right>=8.49~\mathrm{Mrad}$.
The Geant4 particle-resolved scoring gives a mean ionizing energy deposition
of $1.762~\mathrm{keV}$ per incident alpha particle in NbN. Of this energy,
the primary alpha-particle track contributes approximately $52.3\%$, while
secondary electrons contribute the remaining $47.7\%$; contributions from
other particle categories are negligible. For an effective NbN displacement
threshold energy of $E_d=30~\mathrm{eV}$, the corresponding fluence-scaled
NRT displacement damage is $\left<\mathrm{DPA}_{\mathrm{NbN}}\right>=5.672\times10^{-7}$.
The different depth dependences of TID and DPA are consistent with the
underlying stopping physics. Ionizing energy is deposited throughout the
alpha-particle trajectory, with an increase toward the end of range as the
stopping power rises while the helium ion slows down. Nuclear stopping and vacancy production increase more strongly in this region, producing the pronounced
end-of-range peak in displacement damage. Therefore, for alpha particles with MeV energies, the maximum displacement damage is predicted to occur in the sapphire substrate near the alpha-particle end-of-range rather than in the NbN film. This result provides a microscopic explanation for the measured stability of the superconducting and photon-counting properties after irradiation.

\subsection{Superconducting Properties}
\begin{figure}[h!]
    \centering
    \includegraphics[width=\linewidth]{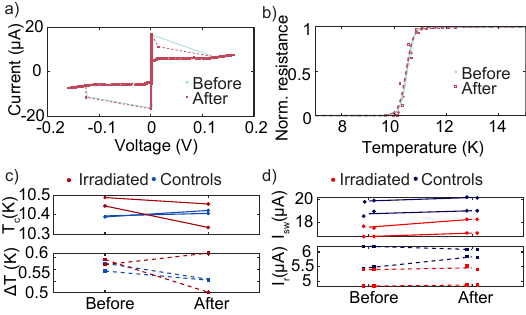}
    %\caption{Electrical characterization before and after exposure. (a) Representative current-voltage characteristics before and after irradiation. (b) Representative normalized resistance-temperature curves before and after irradiation. Changes remain comparable to the variation observed in controls. (c) Superconducting transition temperature and transition width. (d) Switching and retrapping currents. For each condition, the reported values were obtained from separate measurements performed within the same cooldown.}
    \caption{Electrical characterization of the NbN SNSPDs before and after 
the 200-h exposure interval. (a) Representative current-voltage 
characteristics and (b) normalized resistance-temperature curves of an 
irradiated device before and after alpha-particle exposure. 
(c) Superconducting transition temperature, $T_c$, and transition width, 
$\Delta T_c$, for the two control and two irradiated devices before and after exposure. 
(d) Switching current, $I_{\mathrm{sw}}$, and retrapping current, $I_r$, 
before and after exposure. For each of the four devices, two measurements were performed within the pre-exposure cooldown and two within the post-exposure cooldown. Changes in the irradiated devices are comparable to those 
measured in the controls, with no systematic change associated with 
alpha-particle irradiation.}
    \label{fig:electrical}
\end{figure}
Cryogenic electrical characterization shows that the alpha-irradiated devices remain superconducting after exposure with only minor changes to superconducting properties. Representative current-voltage and normalized resistance-versus-temperature curves for an irradiated device are shown in Fig.~\ref{fig:electrical}(a,b), with no large change in the pre- and post-exposure characteristics. Fig.~\ref{fig:electrical}(c) shows the measured transition temperature and transition width before and after exposure. Although the irradiated devices show a small decrease in $T_\mathrm{c}$ after exposure, the change is comparable to the variability observed between devices and between repeated measurements. Similarly, Fig.~\ref{fig:electrical}(d)
shows that the switching and retrapping currents remain within the
range defined by the control devices. The variations measured in the irradiated devices are comparable to those observed in the unirradiated controls. Within the repeatability defined by the control devices, no change in superconducting properties can be attributed to alpha-particle irradiation in this experiment.

\begin{figure}[!t]
     \centering
     \includegraphics[width=0.96\linewidth]{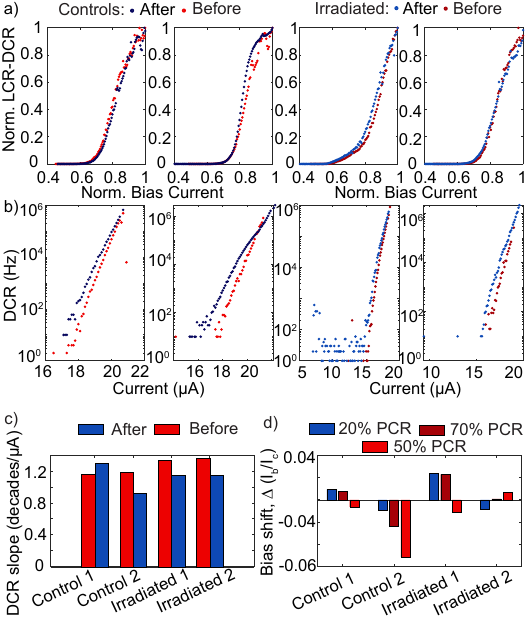}
     %\caption{Photon-counting response before and after exposure. (a) Light-count response, $\mathrm{LCR}-\mathrm{DCR}$, normalized to the maximum value of each bias sweep. (b) Dark count rate before and after exposure. (c) Fitted slope of $\log_{10}(\mathrm{DCR})$ versus bias current before and after exposure. (d) Change in normalized bias current, $\Delta(I_b/I_c)$, required to reach 20\%, 50\%, and 70\% of the normalized photon-count response.}
     \caption{Photon-counting response of control and alpha-irradiated NbN SNSPDs before and after the 200-h exposure. (a) Normalized photon-count response, LCR$-$DCR, versus normalized bias current, $I_b/I_c$. The before/after normalization currents are 19.9/19.4~$\mu$A for Control 1, 18.5/18.4~$\mu$A for Control 2, 17.2/18.1~$\mu$A for Irradiated 1, and 17.8/18.4~$\mu$A for Irradiated 2; each response is normalized to its own maximum. (b) DCR versus bias current. (c) Slopes of $\log_{10}(\mathrm{DCR})$ fitted over $10^3 \leq \mathrm{DCR} \leq 10^5$ counts/s. (d) Shift in normalized bias current required to reach 20\%, 50\%, and 70\% of the maximum photon-count response, defined as $\Delta(I_b/I_c)_p=(I_{b,p}/I_c)_{\mathrm{after}}-(I_{b,p}/I_c)_{\mathrm{before}}$. Irradiated-device changes remain within the range observed for controls.}
     \label{fig:counting}
\end{figure}

\subsection{Photon-Counting Response}

Photon-counting measurements show that the detector response is preserved after irradiation. Fig.~\ref{fig:counting}(a) shows nearly overlapping normalized photon-count curves before and after exposure, indicating preserved bias-dependent response. Fig.~\ref{fig:counting}(b) shows no systematic post-irradiation increase in DCR. The measured changes are comparable to those of the unirradiated controls, suggesting that they are dominated by measurement-to-measurement variations, aging, and device-to-device variability. The quantitative metrics in Fig.~\ref{fig:counting}(c,d) support the same conclusion: the slope of $\log_{10}(\mathrm{DCR})$ versus bias current does not systematically increase, and the bias-current shifts $\Delta(I_b/I_c)$ at 20\%, 50\%, and 70\% of the maximum photon-count response remain small, ranging from approximately $-0.055$ to $0.025$. Overall, the photon-counting data show no measurable irradiation-induced degradation, consistent with the low simulated DPA in the active NbN layer.

\subsection{Radiation-Response Interpretation}

To compare with standard radiation-effects metrics, we calculated the ionizing dose that the same 5.304-MeV alpha-particle fluence would produce in silicon using SRIM ionization profiles. This comparison is not intended as a direct device-equivalence metric, because NbN SNSPDs and silicon electronics have different materials, active volumes, and failure mechanisms. Instead, it provides a benchmark against the conventional krad(Si) scale used in radiation testing. At $\Phi_\alpha=1.27\times10^{12}~\mathrm{cm^{-2}}$, the air-inclusive Geant4 calculation gives an average NbN ionizing dose of 8.49~Mrad, while the corresponding silicon reference calculation gives a dose on the order of tens of Mrad(Si)~\cite{nasa_lesson824}. Thus, the exposure corresponds to a multi-Mrad radiation level on the conventional silicon dose scale~\cite{nasa_lesson824}, yet the NbN SNSPDs show no measurable degradation. This result is further supported by prior 30-keV He$^+$ irradiation studies of related NbN SNSPDs, where selected irradiation conditions preserved or improved detector performance~\cite{incalza2023effects,batson2024effects}. Using the same SRIM-based conversion, those fluences are estimated to correspond to NbN ionizing doses of several hundred megarad, substantially exceeding the dose reached in the present study. The absence of degradation in the present devices is therefore consistent with previous observations and contrasts with semiconductor single-photon detectors~\cite{anisimova2017mitigating}. The absence of degradation is also consistent with the different operating physics of superconducting nanowires and semiconductor single-photon detectors. In SNSPDs, radiation-induced changes would be expected to appear through modifications of the superconducting properties, switching behavior,
or photon-counting response, rather than through the carrier-generation and charge-collection mechanisms relevant to semiconductor detectors
~\cite{wu2022radiation, moscatelli2013radiation, anisimova2017mitigating}.

\section{Conclusion}

We studied the radiation response of NbN SNSPDs using 5.304-MeV alpha-particle irradiation and compared with SRIM and Geant4 radiation-transport simulations. The air-inclusive Geant4 simulations show that the active 5-nm NbN film receives an average fluence-scaled TID of 8.49~Mrad, while the corresponding NRT displacement damage remains small at $5.67\times10^{-7}$ DPA for an effective displacement threshold
energy of 30~eV. This difference arises because the NbN layer is very thin and entirely located in the entrance region of the alpha-particle trajectory, whereas the maximum displacement damage occurs deeper in the sapphire substrate near the alpha-particle end of range.

Cryogenic measurements before and after irradiation show no systematic degradation of the superconducting transition, switching current, retrapping current, dark-count behavior, or photon-counting response. The observed variations are comparable to those measured in unirradiated control devices, indicating that no measurable alpha-irradiation-induced degradation is observed under the tested exposure conditions.

Together, the experimental and simulation results indicate that NbN SNSPDs can tolerate multi-Mrad ionizing doses while preserving superconducting and photon-counting functionality. These results support NbN SNSPDs as radiation-resilient single-photon detectors for space and other radiation-environment applications. Future work should extend this study to additional particle species and energies, including protons, electrons, and heavier ions.

\section*{Acknowledgment}
The authors thank MIT.nano for access to its facilities; J. Daley, C. Moorman, and M. Mondol
for assistance with SEM imaging and electron-beam lithography; and M. S. Galanek for assistance with handling the radioactive source. The authors also thank Phillip D. Keathley, Dongmin Kim, and Davide Mondin for helpful feedback on the manuscript.

\balance
\bibliographystyle{IEEEtran}
\bibliography{references}

\end{document}